# Mid-infrared Electro-Optic Modulation in Few-layer Black Phosphorus

Ruoming Peng[1], Kaveh Khaliji[1], Nathan Youngblood[1], Roberto Grassi[1], Tony Low[1*], and Mo Li[1†]

[1]Department of Electrical and Computer Engineering, University of Minnesota, Minneapolis, MN 55455, USA

**Black phosphorus stands out from the family of two-dimensional materials as a semiconductor with a direct, layer-dependent bandgap in energy corresponding to the spectral range from the visible to the mid-infrared (mid-IR), as well as many other attractive optoelectronic attributes. It is, therefore, a very promising material for various optoelectronic applications, particularly in the important mid-IR range. While mid-IR technology has been advancing rapidly, both photodetection and electro-optic modulation in the mid-IR rely on narrow-band compound semiconductors, which are difficult and expensive to integrate with the ubiquitous silicon photonics. For mid-IR photodetection, black phosphorus has been proven to be a viable alternative. Here, we demonstrate electro-optic modulation of mid-IR absorption in few-layer black phosphorus under field applied by an electrostatic gate. Our experimental and theoretical results find that, within the doping range obtainable in our samples, the quantum confined Franz-Keldysh effect is the dominant mechanism of electro-optic modulation. Spectroscopic study on samples with varying thickness reveals strong layer-dependence in the inter-band transition between different sub-bands. Our results show black phosphorus is a very promising material to realizing efficient mid-IR modulators.**

Recently, two-dimensional (2D) materials, including graphene, various transition metal dichalcogenides, hexagonal boron nitride and black phosphorus (BP), have shown great potential in realizing active optoelectronic components in integrated photonics[1-3]. Those include photodetectors[4-12], optical modulators[13-15], light emitters and laser diodes[16-18]. The diverse optical and electronic properties of 2D materials enable optoelectronic functions covering a broad spectral range from the ultraviolet to the visible, and from the infrared (IR) to the terahertz. Even more exciting is the possibility of designers' optoelectronic materials at desired spectral range with van der Waals heterostructure built from the family of numerous 2D materials[19]. Among all the 2D materials, black phosphorus has the unique property that its bandgap varies with the number of layers over a very wide range, from around 0.3 eV in the bulk to approximately 1.5-2.0 eV in monolayers (the corresponding optical wavelength range is 0.62-4.3 μm)[20-23]. In addition, the hole mobility in BP is as high as $10^4$ cm$^2$/V·s and the effective mass of electrons and holes are small along the armchair crystal direction[24]. All of these make BP a very promising material for broadband, hyperspectral optoelectronic applications. Indeed, many types of photodetectors based on black phosphorus have been demonstrated, operating in the visible and the near-IR bands[10-12,25,26]. These photodetectors show high responsivity and low dark current—thanks to BP's bandgap, which is a significant advantage over graphene—and very high response speed[12]. Although the chemical stability of black phosphorus is a concern as it reacts with water and oxygen and quickly degrades when exposed to air[27,28], BP can be passivated or concealed[29,30], after which the BP devices can be preserved over a very long time without discernible degradation in device performance.

---

[*] Corresponding author: tlow@umn.edu

[†] Corresponding author: moli@umn.edu

Considering multilayer BP, its bandgap size around 0.3 eV makes it ideal for optoelectronic applications in the mid-IR range, which has many important applications such as spectroscopic chemical detection and free-space communication[31,32]. It is highly desirable to realize mid-IR systems that are fully integrated, compact and portable for widespread applications. While mid-IR laser sources have become commercially available[33] and integrated photonic circuits have been developed on the ubiquitous silicon-on-insulator platform[31,34], mid-IR photodetection still relies on narrowband compound semiconductors such as InAsSb and HgCdTe, which are difficult to be integrated with silicon. As an alternative, BP mid-IR photodetectors have already been demonstrated with promising performance[25,26]. For mid-IR optical modulation, multilayer BP is equally promising[35]. Strong field-effect tuning of its bandgap has been observed in BP, and its infrared optical response has been measured[36-38]. It's anisotropic band structure and interband coupling also imbued it with peak-like subband absorption features, which is advantageous for low voltage optical modulation[39-41]. Here we report the broadband measurement of the electrical modulation of the optical absorption in multilayer BP, corroborated with the theoretical calculation of the optical conductivity obtained from Kubo formula based on the tight-binding description[39,40]. Our study reveals that the quantum confined Franz-Keldysh effect is the dominant physical mechanism of electro-optic effect in our samples. Our results confirm the viability of using BP to realize an electro-optic modulator in integrated mid-IR photonic systems[35].

To measure electro-optic modulation in BP, we exfoliated multilayer BP flakes from bulk crystal and used the dry transfer method[42] to transfer them onto a heavily doped silicon substrate with a layer of thermally grown silicon dioxide. To maximize the field at BP/SiO$_2$ interface for optimal mid-IR absorption in BP, we set the SiO$_2$ thickness to be 450 nm based on the calculation of the transfer matrix method. Titanium/gold top electrodes were patterned and deposited immediately after the transfer. Finally, 10 nm of alumina was deposited with atomic layer deposition (ALD) to encapsulate the BP layer and protect it from degradation. Schematic illustration of the device and measurement setup is depicted in Fig. 1a. The optical image of a representative device is shown in Fig. 1b. The thickness of the BP flake in Fig. 1b is determined to be 9 nm with atomic force microscopy (AFM), which corresponds to 17 monolayers. We used a mid-IR optical parametric oscillator (Firefly-IR, M Squared) as the light source outputting in the spectral range of 2.5-3.7 μm. The laser beam is focused on the device with a spot size of 20 μm in diameter. The transmitted light is detected with an InAsSb photodetector as the laser output wavelength is scanned to obtain the transmission spectrum of the device. To measure the electro-optic response, we used the Si substrate as the bottom gate while the top electrodes on the BP are grounded. Because silicon has low absorption in the mid-IR range and the substrate is double-side polished, transmission loss due to absorption by the substrate is negligibly small.

Fig. 2a shows the measured extinction coefficient, defined as $1 - T(V_g)/T_0$, where $T$ ($T_0$) is the transmission through the substrate with (without) BP, when $V_g$ = 0, ±150 V is applied to the back gate. During these measurements, the device is under normal illumination with incident light linearly polarized along the armchair (AC) crystalline direction of the BP. While the overall transmission shows non-monotonic energy dependence due to the interference effect in the multilayers of the device, one can observe the gate induced modulation. To reveal the electrostatic control of the transmission, we define the modulation level $[T(V_g) - T(0)]/T(0) = \Delta T/T(0)$, which is computed from the data in Fig. 2a and plotted in Fig. 2b. The result shows clear dependence on the photon energy with (positive and negative) peaks at 0.38 and 0.43 eV, respectively, where modulation level up to 3% is achieved when the gate voltage is switched between ±150 V. Because BP is linearly dichroic due to its anisotropic optical absorption, we also performed similar measurement with light polarization aligned with the zigzag (ZZ) axis. The result displayed as the dashed line in Fig. 2b, show much weaker modulation level, as expected from theory[35]. The 2D plot in Fig. 2c shows systematic measurement results of the modulation level when both photon energy and gate voltage $V_g$ are continuously scanned, which revealed several features worth noting. First, $\Delta T/T(0)$ remains nearly zero, except in the close vicinity of two characteristic photon energies, $E_a$ =0.38 eV and $E_b$ =0.43 eV, where strong modulation can be observed. Second, the sign of $\Delta T/T(0)$ also shows dependence on the polarity of $V_g$. As $V_g$ changes from negative to positive, at $E_a$, $\Delta T/T(0)$ changes from positive to negative, while at $E_b$, the modulation level experience a negative-to-positive sign flip. In addition, one can also spot in Fig. 2c a third characteristic energy with a relatively weak modulation in close proximity of $E_c$=0.5 eV, where the sign change with gate bias is similar to that observed at $E_a$. Finally, $E_a$ shifts to lower value with increasing $|V_g|$, whereas $E_b$ shifts toward higher value at the same time.

To gain a qualitative understanding of the results, in Fig. 2d, we draw energy band diagrams along the z-axis of BP for three situations of gate bias, $V_g$<0, $V_g$=0, and $V_g$>0. We argue that the characteristic energies ($E_{a,b,c}$) in Fig. 2c, where the modulation extrema take place, nearly coincide with the transition energies, $E_{ij}$, between the $i$-th conduction subband ($c_i$) and $j$-th valence subband ($v_j$). To verify, we compare the experimental values with those obtained with quasi-1D tight binding model:

$$E_{ij} = \Delta_0 - 2(\gamma_c - \gamma_v) + \frac{\pi^2(\gamma_c i^2 - \gamma_v j^2)}{(N+1)^2} \qquad (1)$$

where, $\Delta_0$ is the monolayer bandgap, $N$ is the number of layers, and $\gamma_c$ ($\gamma_v$) denotes the nearest neighbor interlayer coupling for the conduction (valance) bands, respectively. We emphasize that expression (1) is valid for flat band condition only. Although, the latter is not the case in our measurements (to be discussed later), Eq. (1) can still be used to gain an intuition of the characteristic energies origin. Using the values of the parameters given in Ref. [40], the transition energies in the energy range of interest for 9 nm BP can be estimated as $E_{11}$=0.39, $E_{12}$=0.41, $E_{21}$=0.43, $E_{13}$=0.44, and $E_{22}$=0.47 eV. The comparison suggests that the first and third modulation exterma in Fig. 2c are related to same-index transition energies $E_{11}$ and $E_{22}$, respectively. The second characteristic energy in Fig. 2c, however, seems to originate from hybrid transitions, i.e., between sub-bands with $i \neq j$, though the weighted contribution of each hybrid transition is not clear.

We next examine the sign variation in $\Delta T/T(0)$ and its dependence on bias polarity. Hall measurement on the sample has shown that the BP is intrinsically p-doped with a hole concentration of $\rho_h$=5.5×10$^{11}$ cm$^{-2}$ at zero $V_g$. Because of the relatively thick oxide layer of the substrate, $V_g$ applied to the back gate is insufficient to tune the BP to the n-doped regime, though the hole concentration $\rho_h$ decreases as $V_g$ changes from negative to positive. For 11 transition, as $V_g$ becomes more negative, band bending makes the electron and hole wavefunctions more localized toward the opposite sides of the BP, thereby reducing the conduction-valance overlap and its corresponding oscillator strength, as illustrated in Fig. 2d. This leads to a decrease in the absorption at the corresponding photon energy and renders the transmission at negative (positive) $V_g$ to be larger (smaller) than $T(0)$, hence the positive (negative) $\Delta T/T(0)$ signs. For 12 transition, however, the wavefunction overlap increases with the hole concentration. The latter renders a reverse $V_g$ dependence for 12 relative to 11 and therefore, justifies the opposite signs in $\Delta T/T(0)$ at $E_a$ and $E_b$ of Fig. 2c. Similar behavior can also be anticipated for 21, or 13. We emphasize that the observed sign change of the modulation level at $E_b$ due to the hybrid transitions in Fig. 2c suggests that the BP cannot be in the flat-band condition when $V_g$ is zero because the hybrid transitions are forbidden under the flat band condition. Instead, the intrinsic doping may be distributed non-uniformly among the sample layers (see middle panel in Fig. 2d) and cause band bending.

To verify the qualitative picture, we resort to multi-physics numerical techniques to solve for the transmission spectra of the geometry shown in Fig. 1a. The full account of the models employed are described in supplemental material, and the results are summarized in Fig. 3. The calculated transition energy $E_{ij}$ and wavefunction overlap at $S_{ij}$ versus hole density $\rho_h$ are illustrated in Fig. 3a. As shown, in the relevant range of $\rho_h$, only the transitions 11, 12, 21, 22, and 13 lie in the energy range of interest. Among those, 13 has near-zero overlap and therefore does not contribute to BP's absorption. For those transitions that contribute, the overlap decreases with $\rho_h$ for 11 and 22 transitions, while it increases for 12 and 21 transitions. Furthermore, the transition energy $E_{11}$ decreases with $\rho_h$, while the rest are almost not affected by it. The observed trends, namely, the reduction of the optical gap and the localization of the conduction and valance band wavefunctions along the opposite sides of the BP, are indeed consistent with the Franz-Keldysh effect in quantum wells under perpendicular static electric field[43].

The real part of the conductivity due to various subband transitions at three $\rho_h$ values are shown in Fig. 3b. For the 11 transition, one can identify a red-shift in the absorption edge with increasing $\rho_h$, which is also observable for $E_{11}$ in Fig. 3a. Furthermore, the consistent change in both conductivity peaks and overlap wavefunctions with $\rho_h$ suggests a direct link between the two quantities. To compare with the experimental results, we calculate the modulation level $\Delta T/T(0)$ as a function of $\rho_h$ and photon energy as shown in Fig. 3c, where $T(0)$ is the optical transmission when $\rho_h=5.5\times10^{11}$ cm$^{-2}$. From the figure, one can identify three energy intervals, specified as I, II and II, where transmission modulation changes sign with $\rho_h$. Comparing the results in Figs. 3b and c, it is evident that interval I and II are both related to 11 transition. More precisely, interval I originates from the shift in the absorption edge, while II is due to the change in wavefunction overlap. For interval III, however, all four transitions specified in Fig. 3b contribute and the competition between 11 and 22 transitions on the one hand, and 12 and 21 on the other, determines how the modulation changes sign with $\rho_h$. Nonetheless, in general, a negative-to-positive sign flip in $\Delta T/T(0)$ with increasing $\rho_h$ can mainly be attributed to same-index transitions. A positive-to-negative sign change of $\Delta T/T(0)$, however, has a more complex origin as it can be caused by both the energy shift of same-index transitions and the increase in wavefunction overlap of the hybrid transitions. The two mechanisms can be discriminated, however, via inspecting the energy separation of the immediate next sign flip. To elaborate, if the positive-to-negative sign change in $\Delta T/T(0)$ were due to the shift in $E_{ii}$ transition, an opposite sign flip in modulation level should follow at immediate energies. This is due to the fact that for same-index transitions, the red shift in the absorption edge is always followed by a decrease in the wavefunction overlap (see e.g. the top panel in Fig. 3b).

Armed with this general conclusion inferred from Fig. 3, we now re-examine the experimental results in Fig. 2c. Emphasizing again that $\rho_h$ in this sample increases with positive-to-negative $V_g$ change, the negative-to-positive sign flip in $\Delta T/T(0)$ at energies $E_a$ and $E_c$, can be unambiguously attributed to same-index transitions. For the second characteristic energy, $E_b$, the positive-to-negative sign change in $\Delta T/T(0)$ with $\rho_h$ along with the fact that it is well separated from $E_c$ (i.e. the modulation sign change at $E_b$ is not followed by an immediate opposite sign flip), certify that hybrid transitions are indeed responsible. As a final remark, we discuss why the bias induced redshift in 11 transition and its corresponding modulation sign change (i.e. interval I in Fig. 3c) does not appear in the experimental result. As mentioned previously, the edge shift induced sign change should happen in energies just before a sign flip induced by the wavefunction overlap. Since the accessible energy range in our measurements is limited, and the overlap induced sign flip is broadened in the energy range of 0.34-0.4 eV, we might have missed the edge-shift induced sign flip. Alternately, one can argue that the appearance of the edge shift induced sign change is highly sensitive to the losses in BP, which is modeled with a phenomenological loss term, denoted as $\eta$ in the Kubo formula. As shown in Fig. 3b, top panel, with increasing $\eta$, the 11 conductivity peak becomes more broadened, and the absorption edge shift is no longer apparent.

We next focus on how bias-induced modulation of transmission spectra depends on BP thickness. In Fig. 4a, we report the same-index transition energies $E_{ii}$ extracted from the contour plots of modulation level by tracing the modulation extrema. As shown, $E_{11}$ in 9 nm thick device remains nearly unchanged with gate bias. For a different device with 12 nm thick BP, however, we observe $E_{11}$ decrease monotonically for a positive-to-negative change of $V_g$, whereas $E_{22}$ exhibits a maximum at $V_g$=10V. These observations are indeed consistent with the results obtained via our numerical calculation. In Fig. 4b, $E_{11}$ shows a much stronger red shift with increasing $\rho_h$ in thicker BP than in thinner BP. For $E_{22}$, it develops a maximum with $\rho_h$ as the number of layers increases. We should mention that for hybrid transitions, it is not possible to pin down the $V_g$ modulation of individual hybrid transitions as they overlap to form one modulation extrema (e.g. $E_b$ in Fig. 2c).

In Fig. 4d, the same-index transition energies at zero $V_g$ are shown as a function of the number of layers $N$. The strong decrease in transition energies with $N$ is evident. Also shown in the figure is the energy transitions obtained from Eq. 1. While the extracted $E_{22}$ at zero $V_g$ follows Eq. 1, for 11 transition, Eq. 1 cannot properly recover the fast descending trend observed in the measurements. These can be explained by referring to Fig. 2d showing the presence of band bending at zero $V_g$ in our samples. Compared with the 22 transition, the 11 transition is more susceptible to this band bending and therefore does not follow Eq. 1, which describes only the flat band condition. The deviation further increases in thicker samples as the first conduction (valance) subbands move toward lower (higher) energies, thereby sensing the band bending more effectively.

In conclusion, we have demonstrated electro-optic modulation of multilayer BP in the mid-IR range, which is dominated by the quantum confined Franz-Keldysh effect. Peaks of modulation level at different photon energies are related to contributions from optical transitions between different sub-bands, which show strong layer and gate voltage dependence. We expect the performance of a BP electro-optic modulator would be improved significantly when a high-$k$ dielectric material such as $HfO_2$ is used as the gate oxide. When integrated with a waveguide, the light will propagate along the plane of the BP with a much longer interaction length. A recent study, using parameters consistent with our experimental results (see SI), predicts that a modulation level of 0.05 dB/μm is attainable in a waveguide integrated BP modulator. This suggests that a 5 dB modulation depth can be achieved with a 100 μm device[35]. The performance can be further improved with novel device design and integration, as those have been proposed and implemented for graphene-based modulators working in the near-infrared regime[13-15,44]. Above results and prospects indicate that BP is a very promising material in mid-IR optoelectronic devices for a wide range of applications.

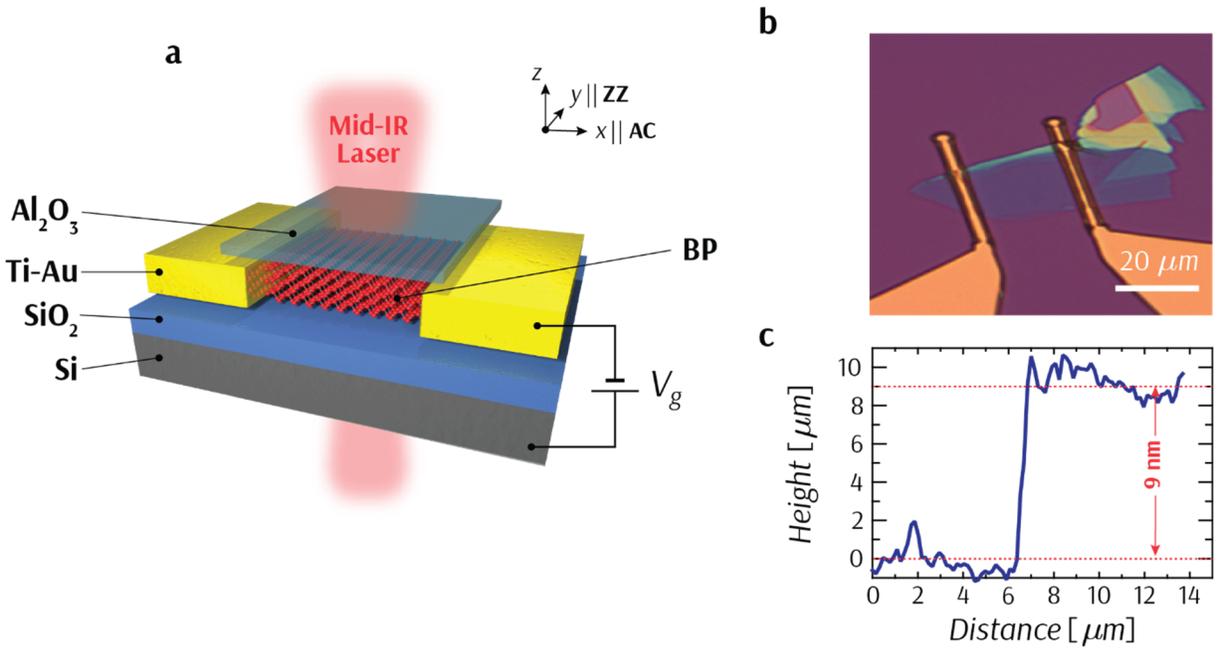

**Figure 1. Black phosphorus electro-optic modulator.** a) Schematic illustration of the BP modulator, featuring the normally incident mid-IR laser beam and the heavily p-doped silicon substrate as the back gate. The BP flake is oriented with the arm-chair (AC) crystalline axis along the x-axis and the zig-zag (ZZ) crystalline axis along the y-axis. b) Optical microscope image of the BP modulator. c) Height profile obtained with atomic force microscope image shows that the BP thickness is about 9 nm which corresponds to 17 layers.

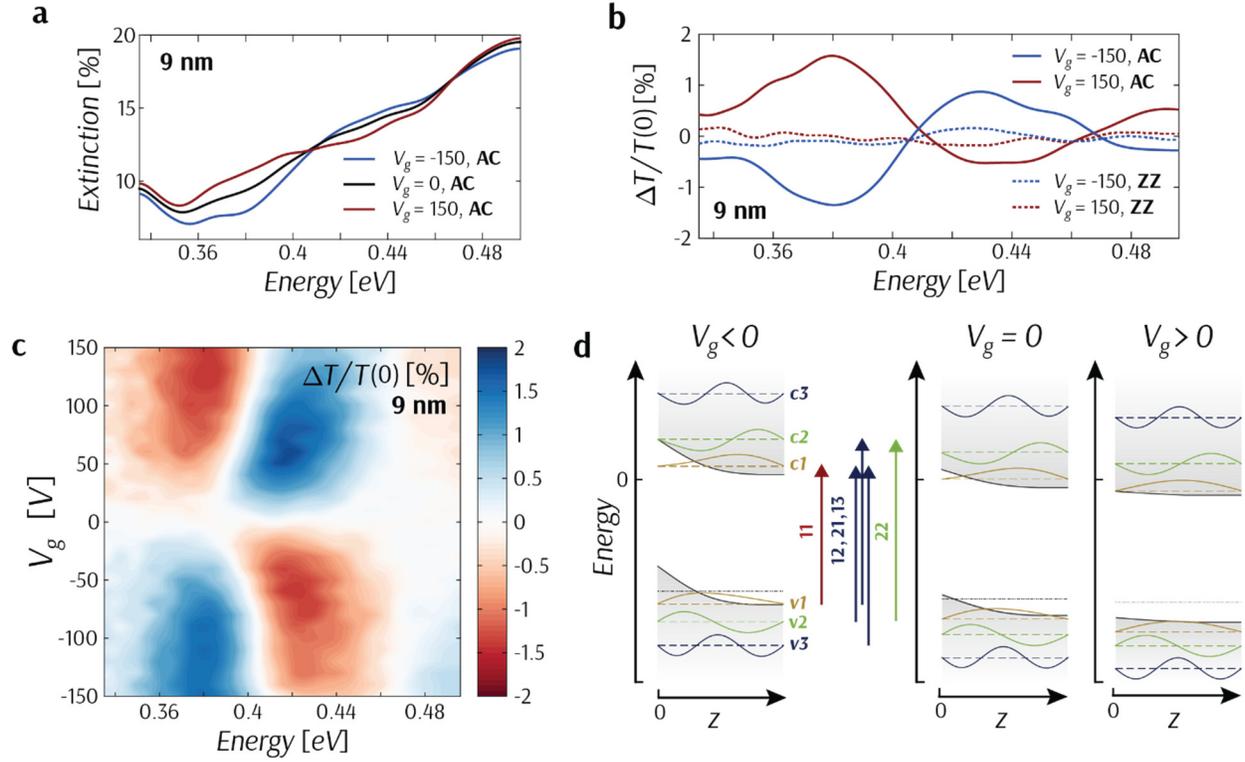

**Figure 2. Measurement of mid-IR electro-optic modulation in BP.** (a) Measured extinction of the transmitted light for 9 nm BP device at three values of gate voltages $V_g = 0, \pm 150$ V. The incident light is linearly polarized with the electric field parallel to the AC crystalline axis. (b) The modulation level $\Delta T/T(0)$ for 9 nm thick BP at $V_g = \pm 150$ V for light polarization along AC (solid lines) and ZZ (dashed lines) crystalline axes. A maximal modulation of about 3% is obtained at a photon energy of 0.38 eV. (c) The modulation level measured as functions of energy and gate bias for 9nm thick BP. Three characteristic peaks can be observed at $E_a = 0.38$ eV, $E_b = 0.43$ eV and $E_c = 0.5$ eV. (d) Schematic energy band diagram of BP under three sample biases, $V_g < 0$, $V_g = 0$, and $V_g > 0$. The dotted lines in (d) mark the Fermi energies. The red/blue/green vertical arrows in (d) show possible transitions that may contribute to the modulation level extrema observed at the characteristic energies of (c).

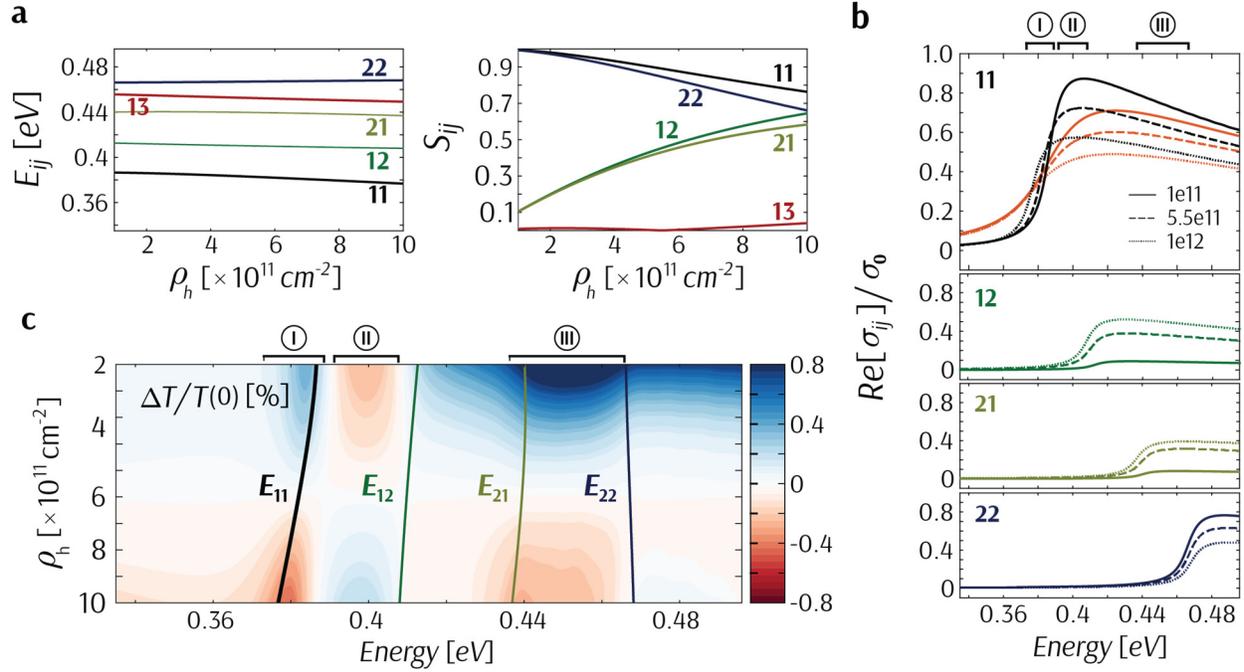

**Figure 3. Theoretical calculation of electro-optic modulation for 9nm thick BP at room temperature.** (a) The transition energy $E_{ij}$ (left panel) and the wavefunction overlap $S_{ij}$ (right panel) as a function of hole concentration $\rho_h$. (b) The real part of conductivity $\sigma_{ij}$ (normalized to $\sigma_0 = e^2/4\hbar$, the universal conductivity of graphene) for various transitions (ij) versus photon energy. (c) Contour plot of the modulation of the transmission as functions of energy and hole concentration. In (b) and (c), broadening is $\eta = 6$ meV, except for the red lines in the top panel (b) where $\eta = 20$ meV.

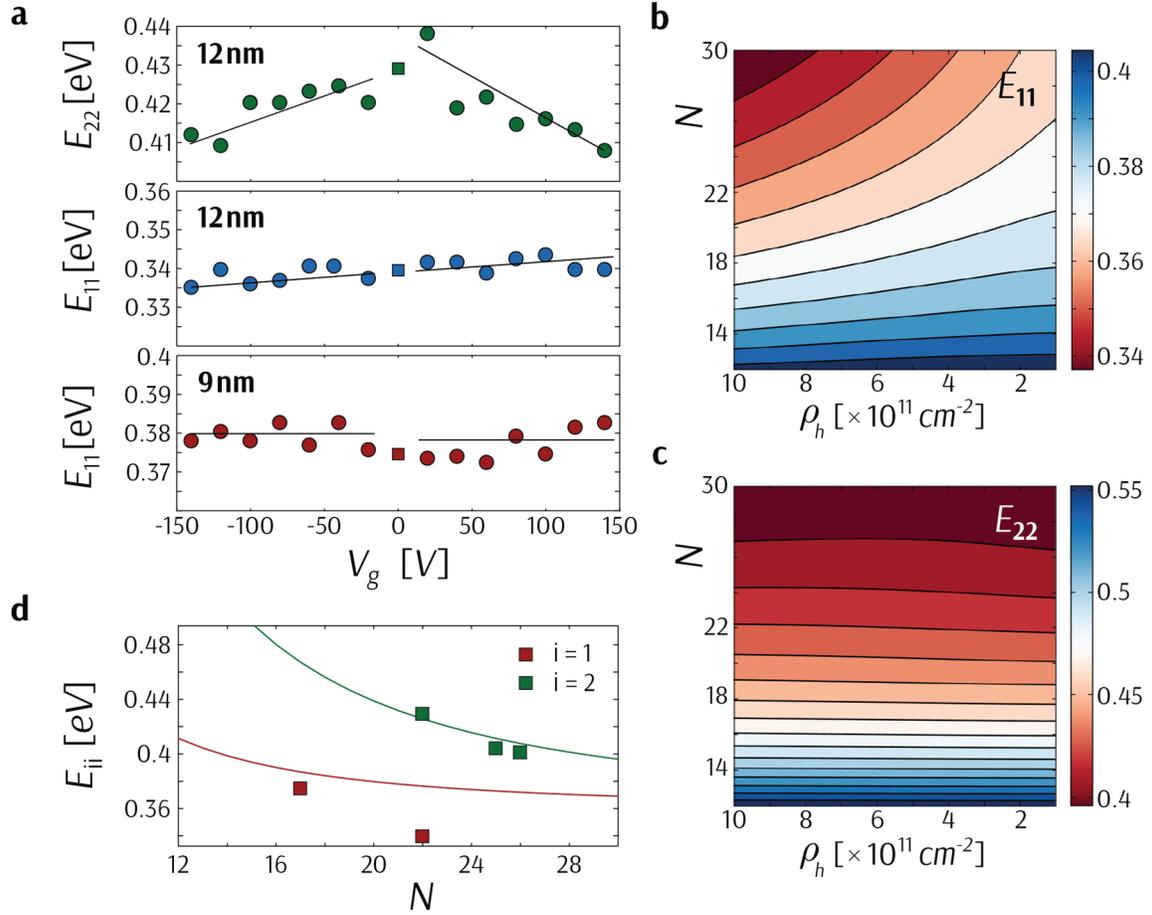

**Figure 4. Layer dependence of electro-optic modulation in BP.** (a) The gate voltage dependence of the transition energy $E_{11}$ and $E_{22}$ in 9 and 12 nm thick BP. Squares denote the transition energy at zero gate voltage, obtained via interpolation of the modulation extrema at $\pm 10$ V. Lines are the guide to the eye. (b, c) Contour plots of $E_{11}$ (b) and $E_{22}$ (c) transitions as functions of BP layer number and hole concentration computed via numerical methods. (d) Measured (symbols) transition energies at zero gate voltage vs. BP layer number. The solid lines are calculated values using Eq. (1).